\documentclass{article}
\usepackage{spconf,amsmath,graphicx}

\title{Projection based advanced motion model for cubic mapping for 360-degree video}
\name{Li Li$^{\ast}$ \qquad Zhu Li$^{\ast}$ \qquad Madhukar Budagavi$^{\star}$ \qquad Houqiang Li$^{\dagger}$}

\address{$^{\ast}$ University of Missouri Kansas City \\
    $^{\star}$  Samsung Research Institute\\
    $^{\dagger}$ University of Science and Technology of China}
\begin{document}
\maketitle
\begin{abstract}
This paper proposes a novel advanced motion model to handle the irregular motion for the cubic map projection of 360-degree video.
Since the irregular motion is mainly caused by the projection from the sphere to the cube map, we first try to project the pixels in both the current picture and reference picture from unfolding cube back to the sphere.
Then through utilizing the characteristic that most of the motions in the sphere are uniform, we can derive the relationship between the motion vectors of various pixels in the unfold cube.
The proposed advanced motion model is implemented in the High Efficiency Video Coding reference software.
Experimental results demonstrate that quite obvious performance improvement can be achieved for the sequences with obvious motions.
\end{abstract}
\section{Introduction}
\label{sec:intro}
When compressing the 360-degree video, it is always needed to project the 360-degree video to 2-D formats to better utilize the commonly used image and video coding standards, such as H.264/Advanced Video Coding \cite{Wiegand2003} and H.265/High Efficiency Video Coding (HEVC) \cite{Sullivan2012}.
Until now, the researchers have already proposed several projection methods \cite{He2016} such as equi-rectangle, cube map, octahedron and so on.
Among them, the cube map is being more and more widely used because it introduces less geometry distortion compared with the equi-rectangle format and simultaneously the cube map is very simple \cite{Zhou2016}.
However, in the cube map, there are also some extent of geometry distortions in all six faces, which may lead to obvious quality degradation under the current translational motion model based motion compensation.
If these geometry distortions can be handled in a suitable way, the coding efficiency of the cube map can be further improved.

The previous approaches, which are proposed to handle complex motions, can be mainly divided into two kinds.
The first kind is the high order motion model based methods, such as affine motion model, bilinear motion model, and perspective motion model.
Among all the high order motion models, the affine motion model is the most frequently considered one.
For example, Wiegend \emph{et al.} \cite{Wiegand2005} propose to use global affine motion model to characterize the global rotation, scaling, and the combination of them.
Besides, Huang \emph{et al.} \cite{Huang2013} introduce a local affine motion model into HEVC to better characterize the local complex motions.
However, such kind of methods is specified for more general complex motions such as rotation and scaling instead of the specific geometry distortions in cube map projection for 360-degree video.
Besides, too many model parameters in the affine motion model may also increase the overhead bits obviously.
  
The second kind is the fish eye motion model based methods to handle the geometry distortions caused by the fish eye cameras.
For example, Jin \emph{et al.} \cite{Jin2015} propose to use the equidistant mapping to characterize the warping in the fish eye lens.
And an efficient motion vector (MV) prediction scheme is proposed to transmit the model parameters efficiently.
Besides, Ahmmed \emph{et al.} \cite{Ahmmed2016} introduce the elastic model into HEVC to better describe the object motions in the fish eye cameras.
Moreover, Eichenseer \emph{et al.} \cite{Eichenseer2016} propose a re-mapping method to handle some extreme cases in which the field of view is larger than 180-degrees.
However, these motion models focusing on the fish eye cameras are unable to efficiently handle the cube map for 360-degree videos.

In this paper, we add a novel advanced motion model into HEVC to handle the geometry distortion in the cube map projection for 360-degree video.
The proposed method first projects the pixels in the unfold cube map to the sphere.
Then through utilizing the characteristic that most of the motions in the sphere are uniform, we can derive the relationship between the MVs among various pixels in the unfold cube.
Besides, we also propose efficient methods to derive more accurate MV predictors for both merge and advanced motion vector prediction (AMVP) modes to further improve the coding efficiency.
In this way, the geometry distortion in the cube map can be handled efficiently and thus the proposed algorithm can provide significant performance improvement.
It should be noted that the proposed method can be easily extended to the other projection formats to handle different extents of geometry distortions.

This paper is organized as follows.
In Section \ref{sec:propose}, we will introduce the proposed advanced motion model in details.
Then the integration of the proposed method with the coding tools in HEVC will be introduced in Section \ref{sec:MV_pred}.
After that, section \ref{sec:experiments} will give the detailed experimental results.
Finally, section \ref{sec:conclusion} concludes the whole paper.

\section{Proposed advanced motion model}
\label{sec:propose}

\begin{figure}[tb]
\begin{minipage}[b]{1.0\linewidth}
  \centering
  \centerline{\includegraphics[width=9.0cm]{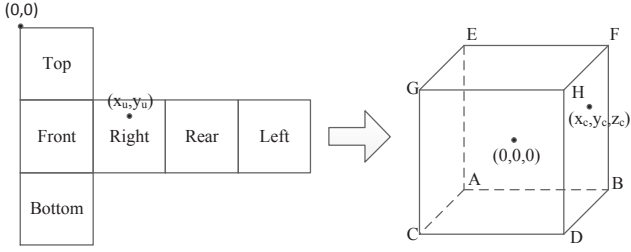}}
\end{minipage}
\caption{Projection from unfold cube map to cube in 3-D space}
\label{fig:2D_3D}
\end{figure}

Since the proposed method tries to characterize the irregular motion in the cube map, we should project both the current pixels and reference pixels in the unfold cube map back to the sphere.
The projection from the unfold cube map to the sphere can be divided into two steps.
Taking the $4\times3$ cube map as an example, we need to first project the unfold cube map to the cube in 3-D space.
The detailed projection process can be seen from Fig. \ref{fig:2D_3D}.
The corresponding coordinate $(x_c,y_c,z_c)$ in the cube of the current coordinate $(x_u, y_u)$ can be calculated through the following formula depending on different faces.
\begin{equation}
\label{eq:2D-3Dx}
x_c=
\left\{ 
\begin{array}
    {c@{\quad}l}
    x_u - faceWidth / 2 & if~face=TOP \\
    x_u - faceWidth / 2 & if~face=FRO \\
    x_u - faceWidth / 2 & if~face=BOT  \\
    faceHeight/2        & if~face=RIG \\
    5 \times faceWidth / 2 - x_u  & if~face=REA   \\
    -faceHeight/2  & if~face=LEF   \\
\end{array}
\right.
\end{equation}
\begin{equation}
\label{eq:2D-3Dy}
y_c=
\left\{ 
\begin{array}
    {c@{\quad}l}
    y_u - faceHeight / 2 & if~face=TOP \\
    faceHeight / 2 & if~face=FRO \\
    5 \times faceHeight / 2 - y_u & if~face=BOT  \\
    3 \times faceWidth / 2 - x_u  & if~face=RIG \\
    -faceHeight / 2 - x_u  & if~face=REA   \\
    x_u - 7 \times faceWidth / 2  & if~face=LEF   \\
\end{array}
\right.
\end{equation}
\begin{equation}
\label{eq:2D-3Dz}
y_c=
\left\{ 
\begin{array}
    {c@{\quad}l}
    faceHeight / 2 & if~face=TOP \\
    3 \times faceHeight / 2 - y_u & if~face=FRO \\
    -faceHeight / 2 & if~face=BOT  \\
    3 \times faceHeight / 2 - y_u  & if~face=RIG \\
    3 \times faceHeight / 2 - y_u  & if~face=REA   \\
    3 \times faceHeight / 2 - y_u  & if~face=LEF  \\
\end{array}
\right.
\end{equation}
It should be noted that in the unfold cube map, the coordinate of the top left pixel is set as $(0,0)$, and in the cube in 3-D space, the coordinate pixel of the center pixel is set as $(0,0,0)$.
The $faceWidth$ and $faceHeight$ are the width and height for each face, respectively.

\begin{figure}[tb]
\begin{minipage}[b]{1.0\linewidth}
  \centering
  \centerline{\includegraphics[width=5.0cm]{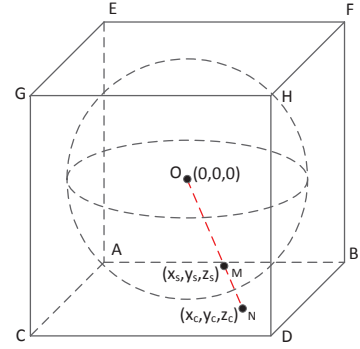}}
\end{minipage}
\caption{Projection from cube in 3-D space to sphere}
\label{fig:3D_Sphere}
\end{figure}

The coordinates $(x_c,y_c,z_c)$ in the cube in 3-D space will be projected to the sphere $(x_s,y_s,z_s)$ as shown in Fig. \ref{fig:3D_Sphere}.
There are two constraints for the point $(x_s,y_s,z_s)$ in the sphere.
One constraint is that $(x_s,y_s,z_s)$ is on the sphere as shown in the following formula.
\begin{equation}
\label{eq:3D-sphere-cos1}
x_s^2 + y_s^2 + z_s^2 = faceWidth^2/4
\end{equation}
The other constraint is that $O$, $M$, and $N$ are on the same line.
\begin{equation}
\label{eq:3D-sphere-cos2}
\frac{x_s}{x_c} = \frac{y_s}{y_c} = \frac{z_s}{z_c}
\end{equation}
Through these two constraints, $(x_s,y_s,z_s)$ can be solved as follows.
\begin{equation}
\label{eq:solutionX}
x_s = \frac{faceWidth/2}{\sqrt{1+\frac{y_c^2}{x_c^2}+\frac{z_c^2}{x_c^2}}}
\end{equation}
\begin{equation}
\label{eq:solutionY}
y_s = y_c \times \frac{x_s}{x_c}
\end{equation}
\begin{equation}
\label{eq:solutionZ}
z_s = z_c \times \frac{x_s}{x_c}
\end{equation}
There are some special cases where $x_c$, $y_c$, or $z_c$ may be $0$.
From Fig. \ref{fig:3D_Sphere}, such cases can be easily handled in a similar way.
In summary, after the above mentioned two steps, the pixel can be projected from the unfold cube map to the sphere.

\begin{figure}[tb]
\begin{minipage}[b]{1.0\linewidth}
  \centering
  \centerline{\includegraphics[width=9.0cm]{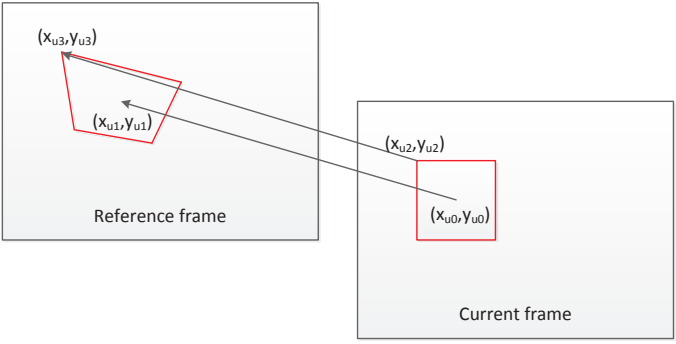}}
\end{minipage}
\caption{MV correspondence for different pixels in a block}
\label{fig:MV}
\end{figure}

As shown in Fig. \ref{fig:MV}, assume that the positions of the center pixel of the current block and the reference block is $(x_{u0},y_{u0})$ and  $(x_{u1},y_{u1})$, respectively.
The problem becomes how to determine the corresponding positions of the other pixels in the reference picture.
Using the pixel $(x_{u2},y_{u2})$ as an example, the main problem is how to derive the value of $(x_{u3},y_{u3})$.

To derive $(x_{u3},y_{u3})$, we first project $(x_{u0},y_{u0})$, $(x_{u1},y_{u1})$, and $(x_{u2},y_{u2})$ in the unfold cube to $(x_{s0},y_{s0},z_{s0})$, $(x_{s1},y_{s1},z_{s1})$, and $(x_{s2},y_{s2},z_{s2})$.
Then utilizing the characteristic that most of the pixels in the sphere are uniform, we can obtain $(x_{s3},y_{s3},z_{s3})$ as follows,
\begin{equation}
\label{eq:uniformX}
x_{s3} = x_{s1} - x_{s0} + x_{s2}
\end{equation}
\begin{equation}
\label{eq:uniformY}
y_{s3} = y_{s1} - y_{s0} + y_{s2}
\end{equation}
\begin{equation}
\label{eq:uniformZ}
z_{s3} = z_{s1} - z_{s0} + z_{s2}
\end{equation}
After obtaining $(x_{s3},y_{s3},z_{s3})$, we can then get $(x_{c3},y_{c3},z_{c3})$ in the cube in 3-D space according to Fig. \ref{fig:3D_Sphere}.
Finally, the $(x_{u3},y_{u3})$ can be calculated through the one-to-one correspondence shown in Fig. \ref{fig:2D_3D}.
The corresponding positions of other pixels in the reference frame can be obtained in a similar way.
After obtaining the corresponding positions of all the pixels in a block, the motion compensation can be performed and so is the following encoding process.
It should be noted that the corresponding positions of some pixels may be in the fractional positions, they are interpolated using the Discrete Cosine Transform based Interpolation Filter (DCTIF) \cite{Lv2012} to up to $1/64$ pixel precision \cite{Lin2015}.
Besides, to determine the initial MV of the center position of the current block, a similar method as Test Zone Search (TZS) algorithm \cite{Purnachand2012}, which is applied to the translational motion model in HEVC, is used to speed up the motion estimation process.

\section{Integration with merge and AMVP mode}
\label{sec:MV_pred}
\begin{figure}[tb]
\begin{minipage}[b]{1.0\linewidth}
  \centering
  \centerline{\includegraphics[width=6.0cm]{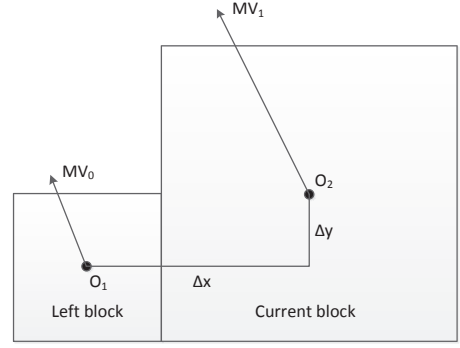}}
\end{minipage}
\caption{MV prediction process}
\label{fig:MV_prediction}
\end{figure}

Besides the advanced motion model based motion compensation process, the MV prediction from the neighboring blocks also has significant influences on the coding efficiency of the proposed framework.
Taking the left block as an example as shown in Fig. \ref{fig:MV_prediction}, assume that the MV of the center of the left block is $MV_0$, then the MV predictor $MV_1$ of the center of the current block can be derived using the same process as in Section \ref{sec:propose} according to their relative distance.
It should be noted that the MV predictor is round to $1/4$ pixel precision to be consistent with the HEVC translational motion model and reduce the overhead bitrate.

For the HEVC translational motion model, two MV prediction schemes including merge and AMVP.
Under the proposed motion model, we also introduce the merge and AMVP schemes into the advanced motion model to better improve the coding efficiency.

\begin{figure}[tb]

\begin{minipage}[b]{0.48\linewidth}
  \centering
  \centerline{\includegraphics[width=4.0cm]{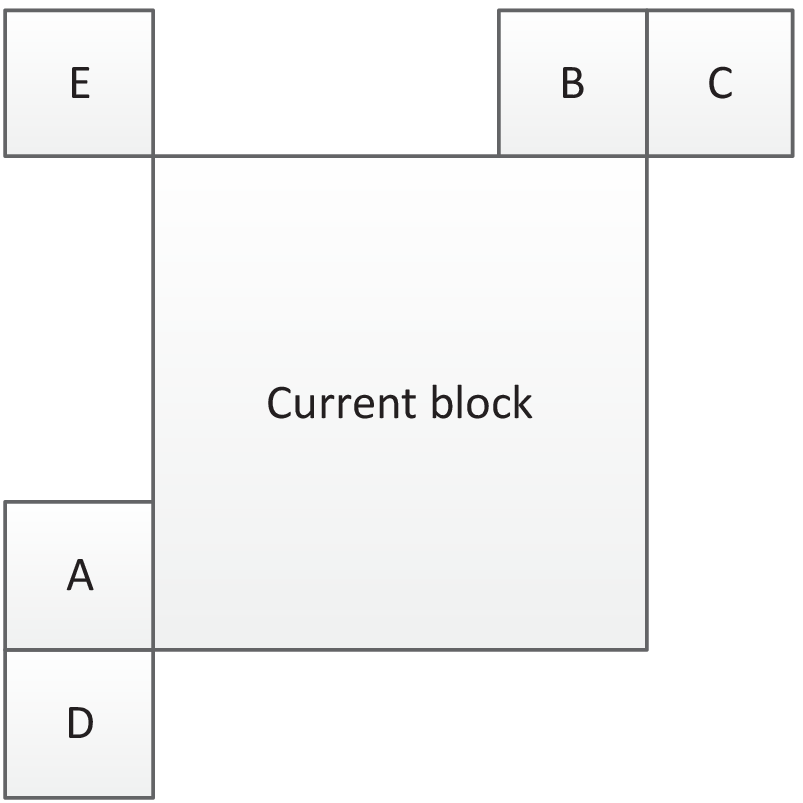}}
  \centerline{(a) Merge}\medskip
\end{minipage}
\hfill
\begin{minipage}[b]{0.48\linewidth}
  \centering
  \centerline{\includegraphics[width=4.0cm]{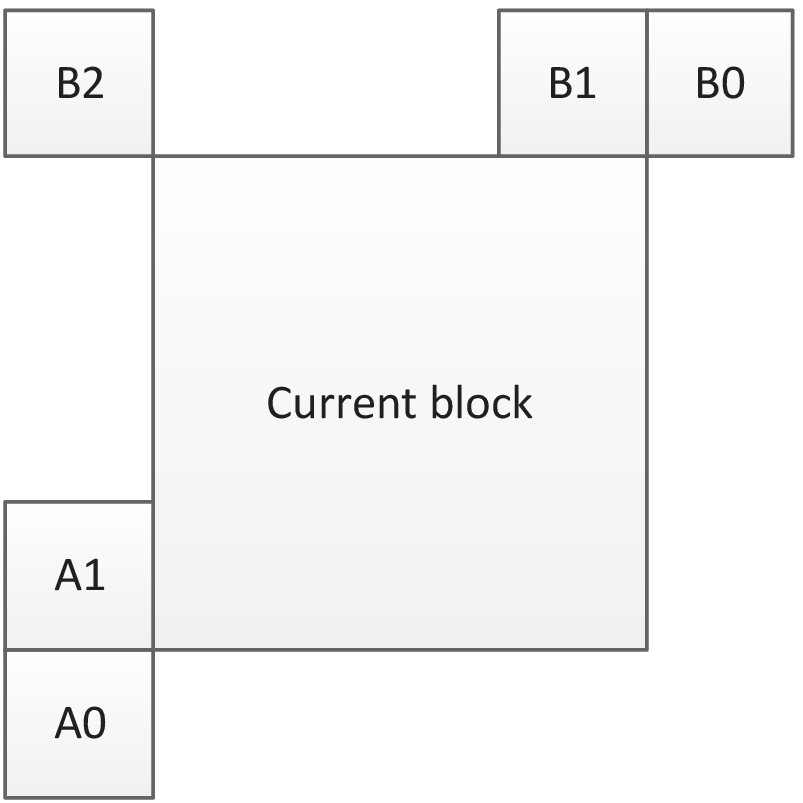}}
  \centerline{(b) AMVP}\medskip
\end{minipage}
\caption{Merge and AMVP candidates}
\label{fig:cand}
\end{figure}

For the merge scheme, as shown in Fig. \ref{fig:cand} (a), similar to the merge mode in translational motion model, we will transverse block A, B, C, D, and E to get the merge candidates.
If no valid merge candidates are found, the advanced motion model is disabled.
The maximum number of merge candidates is set as $1$.
Besides, since when the MV is zero, the proposed motion model and the traditional motion model will be the same, the zero motion is also considered as invalid when searching for a merge candidate.

For the AMVP scheme, as shown in Fig. \ref{fig:cand} (b), we will first search the left block $A0$ and $A1$, and then search the above predictor $B0$, $B1$, and $B2$.
Since in the AMVP scheme, we will perform motion estimation to find the optimal MV, the zero MV is used to as the MV predictor when the MVs of all the neighboring blocks are valid.
The maximum number of AMVP candidates is set as $1$.
Besides, under AMVP mode, the neighboring blocks may point to different reference frames from the given reference frame, in this case, the MV scaling operations are applied.

\section{Experimental results}
\label{sec:experiments}
The proposed algorithm is implemented in HM-16.6 \cite{HM16.6} to compare with the HM16.6 anchor to demonstrate the effectiveness of the proposed algorithm.
We mainly test three configurations including random access (RA) main, low delay (LD) main, and low delay P (LP) main following the HEVC common test condition \cite{Bossen2013}.
The QPs tested are $22$, $27$, $32$, $37$ including high bitrate and low bitrate cases.
Since the proposed algorithm targets at better characterizing the irregular motions in the cube map projection of the 360-degree video, we select some sequences in cubic $4\times3$ formats with relatively larger motion to show the benefits of the proposed algorithm.
The detailed characteristics of the test sequences are shown in Table \ref{tab:seq}.
Besides, since these sequences are all with very high spatial resolutions, only one second is tested to show the benefits of the proposed algorithm.

\begin{table}[t]
\begin{center}
\caption{The characteristics of the test sequences. The first four sequences are converted from the equi-rectangle format. The last two sequences are provided by GoPro. } \label{tab:seq}
\vspace{3pt}
\begin{tabular}{c|c|c}
  \hline
Sequence name & Resolution & frame count    \\
  \hline
  SkateboardInLot     & $4736\times3552$ & 32 \\
  ChairLift           & $4736\times3552$ & 32 \\
  DrivingInCity       & $3840\times2880$ & 32 \\
  DrivingInCountry    & $3840\times2880$ & 32 \\
  Bicyclist           & $3840\times2880$ & 24 \\
  Glacier             & $3840\times2880$ & 24 \\
  \hline
\end{tabular}
\end{center}
\end{table}

\begin{table}[t]
\begin{center}
\caption{The performance of the proposed algorithm RA} \label{tab:RA}
\vspace{3pt}
\begin{tabular}{c|c|c|c}
  \hline
Sequence name & Y & U & V    \\
  \hline
  SkateboardInLot     &  --1.6\% & --1.0\% & --0.9\% \\
  ChairLift           &  --1.3\% & --0.5\% & --0.8\% \\
  DrivingInCity       &  --0.8\% & --0.5\% & --0.8\% \\
  DrivingInCountry    &  --2.6\% & --0.7\% & --1.3\% \\
  Bicyclist           &  --1.2\% & --1.1\% & --1.4\% \\
  Glacier             &  --3.4\% & --2.6\% & --3.0\% \\
  Average             &  --1.8\% & --1.1\% & --1.4\% \\
  \hline
\end{tabular}
\end{center}
\end{table}

\begin{table}[t]
\begin{center}
\caption{The performance of the proposed algorithm LD} \label{tab:LD}
\vspace{3pt}
\begin{tabular}{c|c|c|c}
  \hline
Sequence name & Y & U & V    \\
  \hline
  SkateboardInLot     &  --1.2\% & 0.0\% & --0.8\% \\
  ChairLift           &  --0.5\% & 0.2\% & --0.3\% \\
  DrivingInCity       &  --0.5\% & --0.2\% & --0.8\% \\
  DrivingInCountry    &  --1.2\% & --0.3\% & --0.4\% \\
  Bicyclist           &  --0.9\% & --0.8\% & --0.7\% \\
  Glacier             &  --1.7\% & --0.8\% & --1.2\% \\
  Average             &  --1.0\% & --0.3\% & --0.5\% \\
  \hline
\end{tabular}
\end{center}
\end{table}

\begin{table}[t]
\begin{center}
\caption{The performance of the proposed algorithm LP} \label{tab:LP}
\vspace{3pt}
\begin{tabular}{c|c|c|c}
  \hline
Sequence name & Y & U & V    \\
  \hline
  SkateboardInLot     &  --1.0\% & --0.3\% & --1.3\% \\
  ChairLift           &  --0.4\% & 0.2\%   & --0.4\% \\
  DrivingInCity       &  --0.7\% & 0.0\%   & --0.8\% \\
  DrivingInCountry    &  --1.3\% & --0.5\% & --1.0\% \\
  Bicyclist           &  --1.0\% & --0.3\% & --1.0\% \\
  Glacier             &  --1.8\% & --0.9\% & --1.2\% \\
  Average             &  --1.0\% & --0.3\% & --0.9\% \\
  \hline
\end{tabular}
\end{center}
\end{table}

The experimental results on RA main, LD main, and LP main cases are shown in Table \ref{tab:RA}, Table \ref{tab:LD}, and Table \ref{tab:LP}, respectively.
From these tables, we can see that an average of $1.8\%$, $1.0\%$, $1.0\%$ Bjontegaard Delta-rate (BD-rate) \cite{Bjontegaard2001} savings can be achieved through the proposed algorithm for Y components in RA, LD and LB cases.
Since the average distances between the neighboring frames in encoding order are larger than those in LD and LP cases, the irregular motion will be more obvious in RA case.
That is why we can achieve better R-D performance improvement in RA case.
For U and V components, although we can still achieve some BD-rate reduction, the performance improvement becomes much less compared with the Y component as the contents for the U and V components with much fewer differences.
Especially for the sequence Glacier with relatively larger motion, for the Y component, we can achieve $3.4\%$ R-D performance improvement in RA case.
We believe that the larger the motion of the test sequences, the larger the difference between the advanced motion model and the translational motion model will be.
Therefore, more obvious bitrate savings can be achieved for the sequences with large motions.

\section{Conclusion}
\label{sec:conclusion}
Although the cube map projection can lead to much fewer geometry distortions compared with the equi-rectangular format, there are still some geometry distortions in the cube map projection.
Therefore, this paper proposes a novel advanced motion model to handle the irregular motion for the cubic map projection of 360-degree video.
Since the irregular motion is mainly caused by the projection from the sphere to cube map, we first try to project the pixels in both the current picture and reference picture from unfolding cube back to the sphere.
Then through utilizing the characteristic that most of the motions in the sphere are uniform, we can derive the relationship between the motion vectors of various pixels in the unfold cube.
The proposed advanced motion model is implemented in the High Efficiency Video Coding reference software.
Experimental results show that quite obvious R-D performance improvement can be achieved for the sequences with relatively large motions.
\bibliographystyle{IEEEbib}
\bibliography{ref}

\end{document}